\documentclass[aps,prl,preprint,groupedaddress,showpacs,graphics]{revtex4}

\usepackage{graphicx}
\usepackage{dcolumn}
\usepackage{amsmath}
\usepackage{bm}
\usepackage[usenames]{color}
\begin{document}

\title {Decoherence-Free Emergence of Macroscopic Local Realism for entangled photons in a cavity}
\author{S. Portolan$^1$, O. Di Stefano$^2$, S. Savasta$^2$, F. Rossi$^1$, R. Girlanda$^2$}
\affiliation{$^1$Istituto Nazionale per la Fisica della Materia (INFM) and Dipartimento
di Fisica, Politecnico di Torino, Corso Duca degli Abruzzi 24, 10129 Torino, Italy}
\affiliation{$^2$Dipartimento di Fisica della Materia e Tecnologie Fisiche
Avanzate, Universit\`{a} di Messina Salita Sperone 31, I-98166 Messina, Italy}

\begin{abstract}
{We investigate the influence of environmental noise on
polarization entangled light generated by parametric emission in a
cavity. By adopting a recently developed separability criterion,
we show that: i) self-stimulation may suppress the detrimental
influence of noise on entanglement;
ii) when self-stimulation
becomes effective,
a classical model of parametric emission
incorporating noise provides the same results of quantum theory
for the expectation values involved in the separability criterion.
Moreover we show that, in the macroscopic limit,
it is impossible to observe violations of local
realism with measurements of $n$-particle correlations, whatever n
but finite.
These results provide an interesting example of the emergence of
macroscopic local realism in the presence of strong entanglement even in the absence of decoherence.}

\end{abstract}
\pacs{03.65.Ud, 42.65.Lm}

\maketitle

\newpage
Entanglement is one of the most profound features of quantum
mechanics. It plays an essential role in all branches of quantum
information theory \cite{Nielsen-Zeilinger}. Bell theorem
\cite{Bell}, which is derived from Einstein-Podolsky-Rosen's
(EPR's) notion of local realism \cite{EPR}, quantifies how
measurements on entangled quantum mechanical systems may
invalidate local classical models of reality. While all bipartite
pure entangled states violate some Bell inequality \cite{Gisin},
the relationship between entanglement and non-locality for mixed
quantum states is not well understood yet \cite{Werner,Terhal}. Moreover
recent proposals \cite{EntanglementLaser} and realizations
\cite{Zhang, Macro, Sil, Bowen, Eisenberg} of many-particle
entangled quantum states require a better understanding of the
domain of validity of quantum behaviour. A relevant point is
whether the conflict between classical elements of reality and
quantum mechanics may persist at a macroscopic level \cite{RMPZurek}. In
particular continuous-variable entanglement of intense light
sources has been recently demonstrated in \cite{Zhang,Sil}, and
polarization entanglement of macroscopic beams in \cite{Bowen}.
Moreover, it has been recently shown that a source of strongly
entangled states with photon numbers up to a million seems
achievable \cite{EntanglementLaser}. In these works entanglement
has been tested and quantified by using specific separability
criteria. They consist in inequalities among expectation values of
experimentally measurable quantities, that are violated by
entangled quantum states \cite{Duan}. It is worth noting that
there is a profound conceptual difference between Bell
inequalities and separability (or entanglement) criteria. The
violation of the former implies  quantum features in
conflict with local realism of classical mechanics. Whereas the latter have been
derived assuming from the beginning that we are dealing with a
quantum system.
Specifically demonstrations of these
inseparability criteria (see e.g. \cite{EntanglementLaser,Duan})
exploit the fact that involved operators do not commute.
The behaviour of entanglement
towards a macroscopic situation (even close to classical every-day
life phenomena) and its robustness versus noise and decoherence
are not well understood yet. In this Letter we shall address this
crucial problem focusing on a particular very promising source of
macroscopic entanglement: parametric down-conversion of photons
inside an optical cavity. On the one hand we shall quantify the
detrimental influence of such environment channels and show how
self-stimulation may suppress them efficiently. On the other hand
we shall tackle the problem of the macroscopic limit and of the
emergence of classical elements of reality within a quantum
framework.

In order to investigate the relationship among entanglement,
quantum nonlocality, and the macroscopic limit, we adopt a weaker
nonlocality concept \cite{ReplyZeil} based on an {\em adherence on
the physical system}. In particular we do not ask  quantum
correlations to exceed bounds that cannot be violated by classical
correlations, but we limit us to compare the quantum findings with
the corresponding classical stochastic  calculations for the
specific physical system under investigation. However, a classical
model endowed with stochastic noise represents a realization of a
local stochastic hidden variable theory  and the customary lines
of thoughts are recovered. To this end we consider polarization
entangled light from parametric down-conversion driven by an
intense pump field inside a cavity. The multiphoton states
produced are close approximations to singlet states of two very
large spins \cite{EntanglementLaser}. The interaction Hamiltonian
describing the process is given by
\begin{equation}\label{H}
    \hat H = i \hbar \Omega (\hat a^\dag_h \hat{b}^\dag_v - \hat a^\dag_v \hat{b}^\dag_h) +H.c.\, ,
\end{equation}
where $a$ and $b$ refer to the two conjugate directions along
which the frequency-degenerate photon pairs are emitted. $h$ and
$v$ denote horizontal and vertical polarization and $\hbar \Omega$
is a coupling constant whose magnitude depends on the nonlinear
coefficient of the crystal and on the intensity of the pump pulse.
In the absence of losses, within the Heisenberg picture, the
interaction Hamiltonian in (\ref{H}) dictates the following time
evolution of photon operators:
\begin{eqnarray}
    \hat a_{h,v} = \hat a_{h,v}(0) \cosh (r)  \pm \hat{b}^\dag_{v,h}(0) \sinh (r) \nonumber \\
    \hat b_{v,h} = \hat b_{v,h}(0) \cosh (r) \pm \hat{a}^\dag_{h,v}(0) \sinh (r)\, ,
\label{Heisenberg}\end{eqnarray} where the interaction parameter
$r$ is $\Omega\, \tau$ being  $\tau = L/v$  the interaction time
interval. i.e. the time spent by the photons with velocity $v$
inside a crystal of length $L$. In the absence of losses and
considering the photon vacuum as initial state, the Hamiltonian
(\ref{H}) produces a multiphoton quantum state $\left| \psi
\right>$ that is the polarization equivalent of a spin singlet
state, where the spin components correspond to the Stokes
polarization parameters, $\hat J^A_z = (\hat a^\dag_h \hat a_h -
\hat a^\dag_v \hat a_v)/2$, $\hat J^A_x = (\hat a^\dag_+ \hat a_+
- \hat a^\dag_- \hat a_-)/2$, and $\hat J^A_y = (\hat a^\dag_l
\hat a_l - \hat a^\dag_r \hat a_r)/2$, where $\hat a_{+,-}= (\hat
a_h \pm \hat a_v)/\sqrt{2}$ correspond to linearly polarized light
at $\pm 45^{\circ}$, and $\hat a_{l,r}=(\hat a_h \pm i\hat
a_v)/\sqrt{2}$ to left- and right-ended circularly polarized
light. The label $A$ refers to the $a$ modes. Analogous
expressions can be obtained for $\hat {\bf J}^B$ in terms of the
$b$ modes. It has been shown \cite{EntanglementLaser} that the
state $\left| \psi \right>$ is a singlet state of the total
angular momentum operator $\hat {\bf J} = \hat {\bf J}^A + \hat
{\bf J}^B$. As a consequence   $\left< \right. \psi \left| \right. \hat {\bf J}^2
\left. \right|\psi \left. \right>=0$. Losses, fluctuations and imperfections can
lead to nonzero  values for the total angular momentum,
corresponding to nonperfect correlations between the Stokes
parameters in the $a$ and $b$ beams. Within this picture it is
straightforward to include the presence of noise in the system
assuming that before the pump is switched on the system is in an
incoherent thermal-like state described by a diagonal density
matrix. Dealing with such systems the first analysis one may
perform is an intensity measurement:
\begin{equation}\label{n} n_{ah(v)}(r) \equiv \left< \right. \hat a^\dag_{h,v}\,
\hat a_{h,v}\left. \right> = \sinh^2 r + n_o(1 + 2\sinh^2 r)\, ,
\end{equation}
where $n_o \equiv  \left< \right. \hat{a}^\dag_{h,v}(0)\,
\hat{a}_{h,v}(0)\left. \right> =\left< \right.
\hat{b}^\dag_{h,v}(0)\, \hat{b}_{h,v}(0)\left. \right>$ is the
noise present in the system. It is easy to identify two different
contributions: the first term arises from vacuum fluctuations and
describes true (eventually self-stimulated) spontaneous emission,
vanishing in the absence of parametric interaction. The last term
in Eq.\ (\ref{n}) describes a classical-like amplification of the
input thermal noise $n_0$. It is worth noting that the solution of
the corresponding classical model of the optical parametric
amplifier has the same structure of Eq.\ (\ref{Heisenberg}) with
$a$ and $b$ being of course replaced by classical amplitudes
\cite{Yariv}:
\begin{equation}\label{classicaln} n^C(r) \equiv \left< \right. a^*_{h,v}\,
a_{h,v}\left. \right> =  n^C_o(1 + 2\sinh^2 r)\, ,
\end{equation}
where $\left< \right>$ denotes statistical average and  $n^C_0$
describes, as before, statistical noise. For small $r$ values and
for negligible $n_0$ ($n_0 << r^2$), quantum and classical
descriptions lead to distinct functions of $r$, being $n(r) \simeq
r^2$ and  $n^C(r) \simeq n_C(0)$. This distinct behaviour can be
related to the fact that vacuum fluctuations (in contrast to
classical ones) do not produce work and hence, while they can
stimulate pump scattering, they cannot be directly evidenced by
photodetection. In contrast when $r$ increases, it is no more
possible to distinguish quantum behaviour by means of simple
intensity measurements. In particular for $r \geq 2$ a classical
model with $n^{C}_0 = n_0 +1/2 $ (in order to proper include
vacuum fluctuations), is able to give an intensity description
that cannot be distinguished from the quantum one. It is worth
noting that this behaviour can also be found in intriguing
second-order interference effects \cite{ReplyZeil,MandelPRL} and
it agrees with the old idea that many quanta in a system give rise
to a classical-like behaviour. Other relevant second order
correlations are given by the following anomalous correlators:
\begin{eqnarray}
    A_{hv (vh)}=\left< \right. \hat a_{h(v)}\, \hat b_{v(h)}\left. \right> &=& (n_0 +1/2) \sinh (2r)
    \nonumber\\
    A^C_{hv (vh)}= \left< \right. a_{h(v)}\, b_{v(h)}\left. \right> &=& n^C_0 \sinh (2r)\, ,
\label{Acorr}\end{eqnarray} which quantify the pair correlation
induced by the parametric process. Equation (\ref{Acorr}) shows
that replacing again $n^{C}_0 = n_0 +1/2 $, the classical
description coincides with the quantum one.

If the above  considerations can be formulated for second-order
correlations, now we want to focus our attention on what we can
say about entanglement measurements on this system. In order to
test the degree of entanglement of this correlated quantum system,
a simple inseparability criterion has been derived
\cite{EntanglementLaser}: if $\left< \right. \hat{\bf J}^2\left.
\right>/\left< \right. \hat N \left. \right>$ (where $\hat N= \hat
N_A + \hat N_B$ is the total photon number) is smaller than $1/2$,
then the state is entangled. We now consider the system  at $r
\leq 0$ (before switching on the pump) to be in thermal
equilibrium. In particular we assume the system at $r \leq 0$ to
be in a completely incoherent (mixed) state described by a
diagonal density matrix. The only input for the system are thermal
noise (if $T \neq 0$) and vacuum fluctuations. By using Eq.\
(\ref{Heisenberg})
 and the density matrix for thermal equilibrium, we obtain:
\begin{equation}
    \frac{\left< \right. \hat{\bf J}^2\left. \right>}{\left< \right. \hat N \left. \right>} = \frac{3n_0 (n_0+1)}{4n_0+(1+5n_0)\sinh^2 r}\, .
\label{faseA}\end{equation}
 At zero temperature $n_0=0$ and the system is maximally entangled ($\left< \right. \hat{\bf J}^2\left. \right>=0$)
independently of the magnitude of  $r$. Eq.\ (\ref{faseA}) also
shows that, even when thermal noise  largely exceeds vacuum
fluctuations ($n_0 >> 1$), $\left< \right.\hat {\bf J}^2\left. \right>/\left<
\right. \hat N\left. \right>$ goes below $1/2$ provided that $r$ is sufficiently
large. Moreover $\left< \right. \hat{\bf J}^2\left. \right>/\left<
\right. N \left. \right> \to 0$ for $r \to \infty$.  This result
shows that macroscopic entanglement may in principle be achieved
even in the presence of strong fluctuations, provided that
self-stimulation of the emitted pairs takes place. In particular
the system becomes entangled when $\sinh^2 r > 2 n_0 (3
n_0+1)/(5n_0+1)$. Nevertheless, according to the criterion, in
order to beat  the detrimental effect of strong fluctuations
on entanglement we need to rely on self-stimulation.
Generally speaking it is known that
entanglement as well as violations of Bell inequalities have
limited resistance to noise. In the present  case a small amount
of noise  is sufficient to completely destroy entanglement, e.g.
for $r =10^{-3}$, $n_0=2 \times 10^{-6}$ is sufficient to wash out
entanglement according to Eq.\ (\ref{faseA}); nevertheless it is
sufficient to start self-stimulation (by increasing $r$) to restore it. In order to
get a deeper insight we seek some additional information wondering
if, from the criterion viewpoint (this time), we can distinguish
between classical and quantum findings. To this end we put the two
descriptions (classical and quantum) on equal footing and compute
the entanglement criterion evaluating their differences and
similarities. In particular, a classical calculation, computed
according to the above described prescriptions, gives $\left< {\bf
J}^2\right> / \left< N \right> = 3 n^C_0 / (4+5 \sinh^2 r) $. Of
course classical optics does not require a minimum amount of
fluctuations, thus within a classical model  it is possible to
obtain $\left< {\bf J}^2\right>/\left< N \right> < 0.5$. In the
low excitation regime ($r <<1$) and  $n_0$ lower than $r^2$
classical and quantum calculations of $\left< \right. \hat{\bf
J}^2\left. \right> / \left< \right. \hat N \left. \right>$ display
very different variations with $r$ as it happens for simpler
intensity cases. Moreover the classical calculation above provides
a result that as Eq.\ (\ref{faseA}) goes to zero for $r$ much larger than $n_0$.
This means that experiments
eventually demonstrating macroscopic entanglement for this system
can be accounted for in terms of purely classical correlations,
with no need for a quantum-mechanical explanation. Analogous
conclusions can be reached for different experimentally tested
criteria \cite{Zhang, Sil, Bowen, Eisenberg}. This does not mean
at all that the entanglement criterion is wrong or contradictory,
neither that macroscopic entangled systems cannot display quantum
nonlocality effects. Recently it has been shown that quantum
states of a nondegenerate optical parametric amplifier display
their maximum violation of the Bell inequality due to Clauser,
Horne, Shimony, and Holt just in the macroscopic limit ($r \to
\infty$) \cite{Pan}. However the above analysis suggests that
there is a large class of quantum correlation functions that cannot differ
from the corresponding classical ones in the macroscopic limit.
Indeed we can define  the following set of correlation functions
$\left<\right. \hat {\cal B}^{(n) \dag}_{\alpha}\, \hat {\cal
B}^{(n')}_{\alpha'}\left.\right>$, where $\hat {\cal
B}^{(n)}_{m,l,k} = (\hat b_v)^{n -m}(\hat b_h)^{m-l} (\hat
a_v)^{l-k}(\hat a_h)^k$ is a generic $n$-particle destruction
operator. These correlation functions (and also their classical
counterparts $\left< \right.  {\cal B}^{(n)*}_{\alpha}\,  {\cal
B}^{(n')}_{\alpha'}\left. \right>$) are different from zero only
if $n=n'$ and $\alpha \equiv (k,l,m) = \alpha'$. Since we are
dealing with a Gaussian system \cite{GardinerZoller}, such
correlators (as well as their classical counterparts) can be
decomposed in a sum of products of second order correlation
functions only ($n(r)$ and $A(r)$). Since $A(r)= A^C(r)$ (for
$n^{C}_0 = n_0 +1/2 $), and $n(r)/n^C(r) \to 1$ for $r \to
\infty$, we obtain that
\begin{equation}
\lim_{r \to \infty} \frac{\left< \right. \hat {\cal B}^{(n)
\dag}_{\alpha} \hat {\cal B}^{(n)}_{\alpha}\left. \right>} {\left<
\right.  {\cal B}^{(n)*}_{\alpha}\,  {\cal B}^{(n)}_{\alpha}\left.
\right>} = 1\, . \label{Lim}\end{equation} This result implies
that {\em it is impossible to observe violations of macroscopic
local realism (e.g. violations of Bell inequalities, including
those which are not yet known) by measurements of any finite set
of expectation values that can be expanded as a finite sum of
these correlation functions}. It is easy to verify via explicit
calculations that convergence of limit (\ref{Lim}) is very fast
even for large values of $n$. Bell inequalities can be
schematically expressed as $\mathcal{F}(\left<\right. {\cal
B}^{(n) *}_{\alpha}\,  {\cal B}^{(n)}_{\alpha}\left.\right>(r))
\leq \mathcal{L}(n)$, where $\mathcal{L}(n)$ is a bound imposed by
local realism that cannot be violated by classical correlations,
and $\mathcal{F}$ is a  generic continuous function of
$\left<\right. {\cal B}^{(n) *}_{\alpha}\,  {\cal
B}^{(n)}_{\alpha}\left.\right>$ depending also on the different
settings choosen by the observers.   From Eq.\ (\ref{Lim}):
$\mathcal{F}(\left<\right. \hat {\cal B}^{(n) \dag}_{\alpha}\,
\hat {\cal B}^{(n')}_{\alpha'}\left.\right>(r)) \to
\mathcal{F}(\left<\right. {\cal B}^{(n) *}_{\alpha}\,
{\cal B}^{(n')}_{\alpha'}\left.\right>(r))$ when $r \to \infty$,
thus
any bound ${\cal L}$ cannot be violated (in the limit). One
example of these wide class of Bell inequalities can be found in
\cite{Munro}. These results can be generalized to multipartite
situations obtained e.g. by inserting in the setup a number of
beamsplitters \cite{Zuk}.

So far we have considered a situation where the system is
initially in thermal equilibrium, but eq.\, (\ref{Heisenberg}) has
been obtained under the hypothesis that the system has no losses,
hence it is assumed that for the steady-state calculations at any
value of the interaction parameter $r$ the system is disconnected
from the environment. However in real systems amplification,
losses, and noise disturbances actually happen simultaneously. In
the presence of losses, the (steady-state) Heisenberg equations
have to be replaced by (t-dependent) Langevin equations with noise
sources. In the symmetric case (equal losses for all the four
modes), we obtain,
\begin{eqnarray}
    \hat a_{h}(t) &=& \hat a_{h}(0) \cosh \Delta(t,0) + \hat b^\dagger_{v}(0) \sinh
    \Delta(t,0)\nonumber \\ &+& \int_0^t e^{-\lambda (t-t')} \cosh \Delta(t,t')
    \hat f_{a h}(t')\, dt' \nonumber \\ &+& \int_0^t e^{-\lambda (t-t')} \sinh \Delta(t,t')
    \hat f^\dag_{b v}(t')\, dt'\end{eqnarray}
where $\Delta(t,t')= \int_{t'}^t \kappa(t'')\, dt'' =
\frac{\kappa_0}{\Lambda} (e^{- \Lambda t'}-e^{- \Lambda t})$;
$\hat f_{\alpha}(t)$ are Bose quantum noise operators associated
with the losses \cite{ScullyBook} ($\alpha$ denotes the specific
mode e.g. $\alpha \equiv (a,h)$). In the following we will assume
$\left< \right. \hat f^\dag_{\alpha}(t) \hat f_{\alpha'}(t')
\left. \right>= n_0\,  \delta_{\alpha,\alpha'}\, \delta(t-t')$.
Analogous expressions can be obtained for the other three modes.
\begin{figure}[h!]
\resizebox{!}{6.0 cm}{
\includegraphics{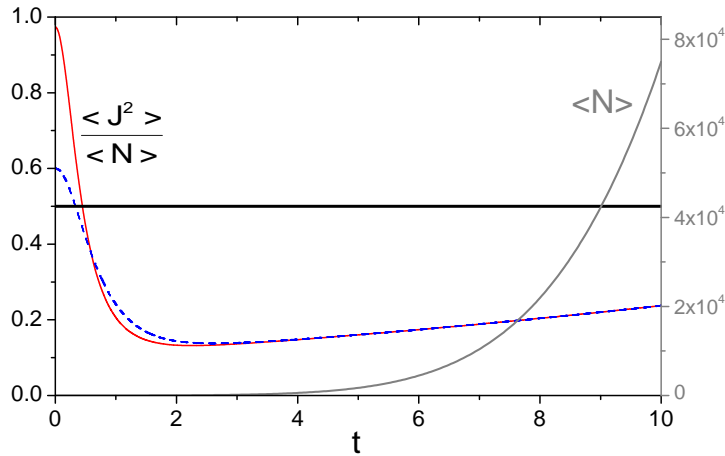}}
\caption{(color online) Time development of $\left< \right. {\bf
J}^2\left. \right> / \left< \right. N \left. \right>$ according to
classical (dashed line) and quantum (continuous line) mechanics,
and quantum calculation of  of the total mean photon number
$\left< \right. N \left. \right>$. Parameters are given in the
text.
}
\label{separabCrit02}
\end{figure}
Fig. \ref{separabCrit02} displays the quantum and the classical
calculation of $\left< \right. \hat {\bf J}^2\left. \right> /
\left< \right. N \left. \right>$ for $\lambda = \Lambda = 0.1$,
$\kappa_0 = 1$. For the quantum (continuous line) and the
classical (dashed line) calculations we adopted $n_0 =0.3$ and
$n^{(cl)}_0 =0.8$ respectively. The figure clearly shows that for
$r > 3$ classical and quantum results cannot be distinguished.
These results show that i) also in this more realistic case
self-stimulation can suppress the detrimental effect of noise ii)
coincidence between classical and quantum results can be obtained
without requiring $n>>1$ if we choose  $n^{(cl)}_0= n_0 +1/2$;
iii) the decoherence due to losses and noise seems to affect
equally quantum entanglement and classical correlations, hence it
cannot be invoked in the present case for the emergence of a
classical behaviour. In order to interpret our results, we
distinguish between two situations: When $r <<1$, the probability
that the system gives rise to states with more than two-photons
 is negligible, so measurements of  $\left< \right. {\bf \hat J}^2\left. \right>$ and of $\left< \right. \hat N\left. \right>$
 can probe the system at a microscopic level, but with a lot of particles (when $r$
increases) both  $\left< \right. {\bf \hat J}^2 \left. \right>$
and $\left< \right. \hat N \left. \right>$ become  macroscopic
observables unable to probe the system fluctuations with precision
at a few quanta level. In this case the information recovered by
observations  is a coarse grained quantity missing the underlying
quantum structure. This lack of microscopic information seems able
to introduce elements of local realism even in the presence of
strong entanglement and in the absence of decoherence. Of course the lack of information
here described should not be confused with a lack of precision of meausrements,
here assumed with unlimited precision.
As shown in
Ref.\ \cite{Zanardi}, the partition of a quantum system into
subsystems and hence the entanglement structure, is dictated by
the set of operationally accessible interactions and measurements.
A given set can hide a multipartite structure. Our results suggest
that, analogously, the set of operationally accessible
measurements and their ability to catch the quantized structure of
the system, determines the quantum or classical nature  of the
observed correlations. As pointed out above, our findings
(included Eq. (\ref{Lim})) do not imply that macroscopic entangled
systems cannot display quantum nonlocality effects. As already
mentioned, it has been shown recently that these kind of systems
do violate CHSH Bell inequality just in the microscopic limit ($r
\to \infty$) and hence do display nonlocality features. However in
that case  the Bell operator is constructed by means of operators
with single quantum sensitivity independently of the number of
particles in the system in contrast to operators $\hat {\cal
B}^{(n) \dag}_{\alpha} \hat {\cal B}^{(n)}_{\alpha}$. Of course
these operators cannot be expanded in a finite sum of operators
$\hat {\cal B}^{(n) \dag}_{\alpha} \hat {\cal B}^{(n)}_{\alpha}$.

The example of the emergence of macroscopic local
realism in the presence of strong entanglement, shown here,
provides insight into the boundary between the classical and
quantum world. These results, with the care that they have been
obtained for a Gaussian system, despite the feasible realization
of systems with a huge amount of entangled particles, suggest that
the lack of information gathered by coarse-grained observations
may lead to the introduction of elements of local realism even in
the presence of strong entanglement and in the absence of
decoherence. In particular, Eq.\ (\ref{Lim}) shows  that
violations of local realism for macroscopic Gaussian states  are
impossible with an apparatus able to measure finite-order
correlations. Further investigations are needed to understand if
the results here presented for  entangled Gaussian systems can be
extended to more general quantum systems.
\begin{acknowledgements}
We thank P.\ Zanardi for helpful discussions and  comments.
\end{acknowledgements}


\end{document}